\documentclass{article}
\usepackage{spconf,amsmath,graphicx}
\usepackage[T1]{fontenc}

\title{A joint separation-classification model for sound event detection of weakly labelled data}
\name{Qiuqiang Kong*, Yong Xu*, Wenwu Wang, Mark D. Plumbley\thanks{* The authors contribute equally to this work.}}
\address{Center for Vision, Speech and Signal Processing, University of Surrey, UK}
\email{\{q.kong, yong.xu, w.wang, m.plumbley\}@surrey.ac.uk}

\begin{document}
\maketitle
\begin{abstract}
Source separation (SS) aims to separate individual sources from an audio recording. Sound event detection (SED) aims to detect sound events from an audio recording. We propose a joint separation-classification (JSC) model trained only on weakly labelled audio data, that is, only the tags of an audio recording are known but the time of the events are unknown. First, we propose a separation mapping from the time-frequency (T-F) representation of an audio to the T-F segmentation masks of the audio events. Second, a classification mapping is built from each T-F segmentation mask to the presence probability of each audio event. In the source separation stage, sources of audio events and time of sound events can be obtained from the T-F segmentation masks. The proposed method achieves an equal error rate (EER) of 0.14 in SED, outperforming deep neural network baseline of 0.29. Source separation SDR of 8.08 dB is obtained by using global weighted rank pooling (GWRP) as probability mapping, outperforming the global max pooling (GMP) based probability mapping giving SDR at 0.03 dB. Source code of our work is published.
\end{abstract}
\section{Introduction}
Sound event detection (SED) aims to detect specific audio events from an audio recording. SED has many applications in our daily life, for example, detecting a baby cry at home, detecting the tapping sound in an office and monitoring the fire alarm or gunshot \cite{valenzise2007scream} in a public area. On the other hand, source separation (SS) aims to separate individual sources from a recording \cite{huang2014deep} and can be used in SED \cite{heittola2011sound}. 

Many current SED models are trained using supervised learning methods \cite{mesaros2016tut, stowell2015detection, heittola2013context}. These supervised learning methods need labelled onset and offset time of the audio events, which we call \textit{strongly labelled data} (SLD). Labelling the SLD is time consuming and difficult to scale \cite{mesaros2016tut}. In addition, the onset and offset time of some audio events are ambiguous due to the fade in and fade out effect, for example, the approaching and moving away of a car. In contrast to the SLD, many audio datasets contain only the tags, that is, the presence or absence of audio events in an audio recordings. This is referred to as \textit{weakly labelled data} (WLD) \cite{kumar2016audio}. Many audio tagging datasets are weakly labelled \cite{foster2015chime, gemmeke2017audio, mesaros2017dcase} and are often larger than strongly labelled SED datasets \cite{mesaros2016tut, gemmeke2017audio}. To utilize the WLD, some methods including joint detection-classification (JDC) model \cite{kong2017joint}, attention and localization model \cite{xu2017attention} and multi-instance learning methods \cite{kumar2016audio} have been used. Source separation can be used for sound event detection \cite{heittola2011sound}. Unsupervised source separation methods including computation audio scene analysis (CASA) uses time-frequency (T-F) masking to emulate how human performs source separation \cite{wang2006computational}. Supervised source separation methods need clean sources for training \cite{huang2014deep} and have achieved state-of-the-art performance. 

In this paper, a \textit{joint separation-classification} (JSC) model is proposed to train the source separation model on the WLD. The proposed framework consists of two parts. The first part is a \textit{separation mapping} from the T-F representation of an audio signal to the T-F segmentation masks of each audio event. The second part is a \textit{classification mapping} from each segmentation mask to its corresponding audio tag. In the source separation stage, separated sources of different classes can be obtained from the T-F segmentation masks. 

The remainder of the paper is organized as follows: Section 2 discusses convolutional neural network. Section 3 proposes the source separation framework. Section 4 shows experimental results. Section 5 concludes and proposes the future work. 

\begin{figure}[t]
  \centering
  \centerline{\includegraphics[width=\columnwidth]{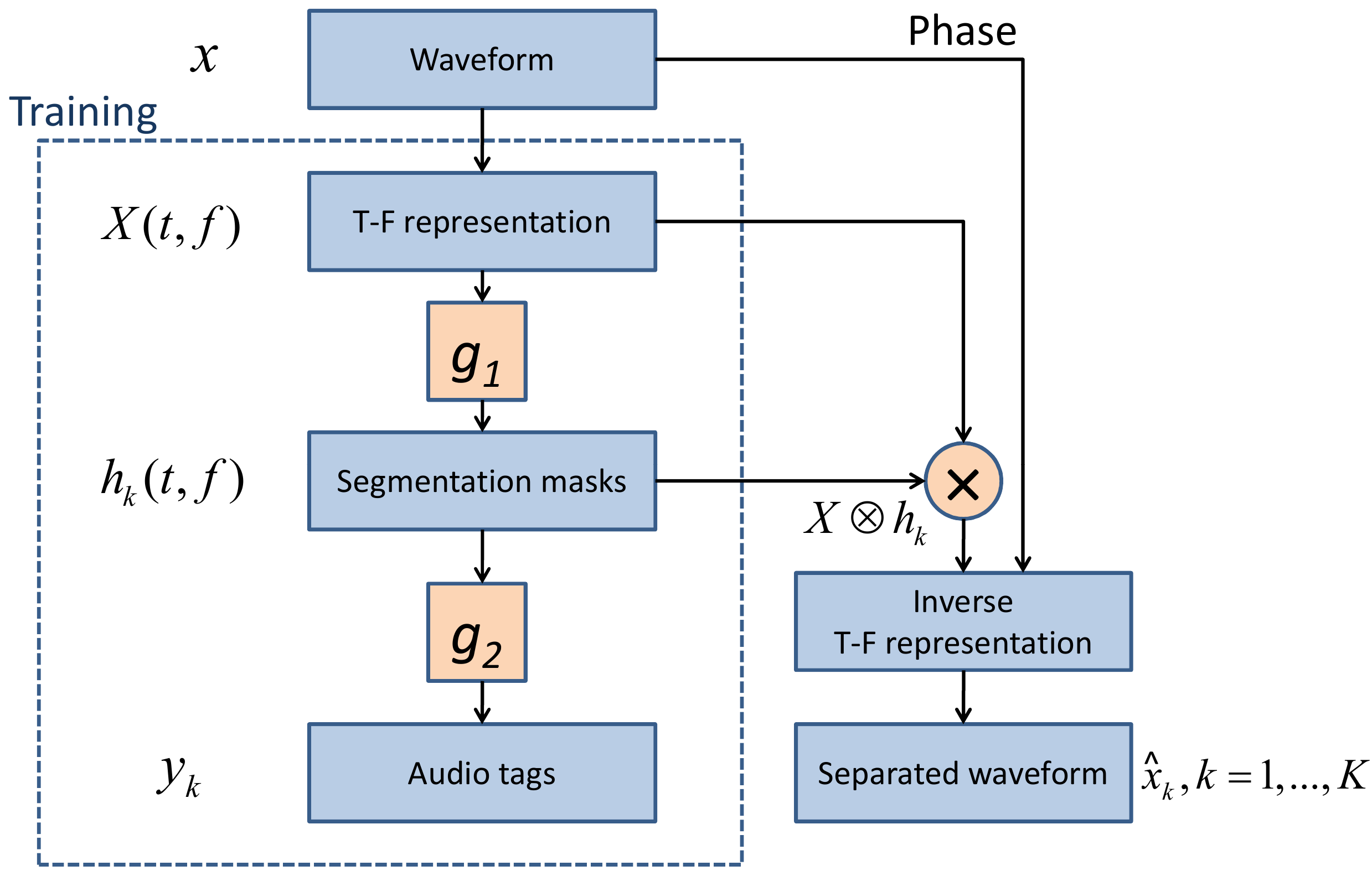}}
  \caption{Framework of the joint separation-classification model. }
  \label{fig:flowchart}
\end{figure}

\section{Convolutional neural network}
\begin{figure*}[t]
  \centering
  \centerline{\includegraphics[width=\textwidth]{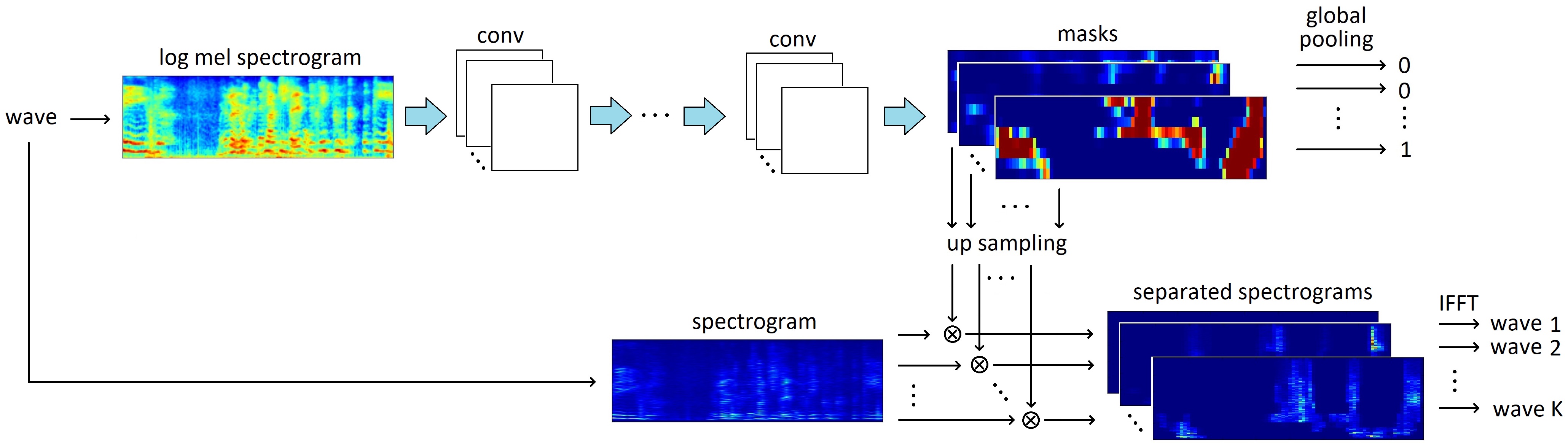}}
  \caption{Convolutional neural network (CNN) based weak source separation. Log Mel spectrogram is used as T-F representation. Separation mapping is modeled by a CNN. Classification mapping is applied on each T-F segmentation mask to obtain the prediction of audio tags. In separation stage, the separated waveforms are obtained from the segmentation masks. }
  \label{fig:CNN}
\end{figure*}

Convolutional neural networks (CNNs) are used initially in image classification \cite{simonyan2014very} and recently have been very successful in audio processing, including speech recognition and audio classification \cite{choi2016automatic}. In audio classification, the waveform is transformed to T-F representations which are treated as an image and fed as input to a CNN \cite{choi2016automatic}. A CNN consists of several convolutional layers and each contains several trainable filters trained to learn some local patterns in the feature map. Downsampling usually follows some convolutional layers to reduce the size of the feature maps. Finally a global max pooling on each feature map \cite{choi2016automatic} is usually used to select the most prominent T-F unit in each feature map followed by a fully connected neural network for classification \cite{choi2016automatic}. 

\section{Proposed joint separation-classification model}
In this section, a joint separation-classification (JSC) model trained on WLD is proposed. This idea is related to the object localization from weakly labelled images \cite{zhou2016learning, oquab2015object}, where only the labels of an image are known, but the location of the objects are unknown. In \cite{zhou2016learning} a class activation mapping (CAM) is applied to highlight the class-specific discriminative regions to localize objects from weakly labelled data. 

Similar to the weakly labelled image data \cite{zhou2016learning, oquab2015object}, many audio datasets \cite{mesaros2016tut} only contain the tags of an audio recording, but the happening time of the events are unknown. The proposed separation-classification model is shown in Fig. 1. The input audio waveform $ x $ is transformed to a time-frequency (T-F) representation $ X(t, f) $ such as a spectrogram or log Mel spectrogram. To simplify the notation we abbreviate $ X(t, f) $ as $ X $. The first part of the model is a \textit{separation mapping} $ g_{1}: X \mapsto \mathbf{h} $ from the input T-F representation $ X $ to the T-F segmentation masks $ \mathbf{h}=[h_{1}, ..., h_{K}] $, where $ K $ is the number of audio tags and $ h_{k} $ is the T-F segmentation mask of the $ k $-th audio tag. The values on each segmentation mask are between 0 and 1 for source separation. The mapping $ g_{1} $ can be parametrized by trainable parameters. The second part of the model is a \textit{classification mapping} $ g_{2}: h_{k} \mapsto y_{k}, k=1, ..., K $ from each segmentation mask to its corresponding audio tag, where $ y_{k} \in [0,1] $ represents the presence probability of the $ k $-th event in an audio recording. A compound model $ g_{2} \circ g_{1} $ is a mapping from the input T-F representation $ X $ to the audio tags $ y_{k}, k=1, ..., K $. In the training phase, the model can be trained end-to-end from $ X $ to $ y_{k}, k=1, ..., K $. 
In the separation stage, the T-F representation of an input waveform is passed through the mapping $ g_{1} $ to get the segmentation masks. Then the input T-F representation is multiplied by each segmentation mask to obtain the separated T-F representation of each event with corresponding audio tag. Then an inverse T-F transform is applied on each separated T-F representation of each audio tag using the phase of the original waveform to obtain its separated waveform of each audio tag (Fig. 1). Finally, SED result of each audio event can be obtained from its corresponding segmentation masks. 

\subsection{Separation mapping}
We apply log Mel spectrogram as input T-F representation, which is a good representation for audio tagging \cite{choi2016automatic}. 
We apply a CNN to model the separation mapping $ g_{1} $. The CNN modeled JSC model is shown in Fig 2. We remove all the downsampling layers to keep the resolution of each T-F segmentation mask the same as the input T-F representation. This high resolution T-F segmentation mask is useful for source separation. The number of feature maps in the last convolutaional layer is the same as the number of audio events to separate followed. Then a sigmoid nonlinearity is applied on the feature maps to obtain the segmentation to ensure the values on segmentation masks are between 0 and 1. This segmentation mask of this T-F representation is similar to the class activation mapping (CAM) in weak image localization \cite{zhou2016learning}.

\subsection{Classification mapping}
The classification mapping $ g_{2} $ maps each segmentation mask to the presence probability of its corresponding tag. Classification mapping can be modeled by, for example, global max pooling (GMP) \cite{zhou2016learning}, global average pooling (GAP) \cite{lin2013network, zhou2016learning} or global weighted rank pooling (GWRP) \cite{kolesnikov2016seed}.

\subsubsection{Global max pooling}
Global max pooling (GMP) \cite{choi2016automatic} is defined as follows:

\begin{equation} \label{eq1}
g(h_{c}) = \underset{t,f}{\max} ~ h_{c}(t,f)
\end{equation}

\noindent where $ h_{c} $ represents the $ c $-th segmentation mask and $ t $, $ f $ are indexes of time and frequency bin. GMP returns the highest value on each feature map. GMP performs well in classification but tends to underestimate the T-F units of events in each segmentation mask \cite{zhou2016learning} because only the T-F unit with the highest value is passed to the next layer (Fig. 3).

\subsubsection{Global average pooling}
Global average pooling (GAP) \cite{lin2013network} is defined as:

\begin{equation} \label{eq1}
g(h_{c}) = \frac{1}{M} \sum_{t,f}^{} h_{c}(t,f)
\end{equation}

\noindent where $ M $ is the number of time frames multiplied number of frequency bins. In contrast to GMP, GAP averages all the values of T-F units on a segmentation mask, which tends to overestimate the events in a segmentation mask \cite{kolesnikov2016seed} (Fig. 3). 

\subsubsection{Global weighted rank pooling}
Global weighted rank pooling (GWRP) is proposed in \cite{kolesnikov2016seed} and is a generalization of GMP and GAP. Define $ I^{c} = \{i_{1}, ... i_{n}\} $ as an index set in descending order of the values on feature map $ h_{c} $, i.e. $ (h_{c})_{i_{1}} \geq (h_{c})_{i_{2}} \geq ... \geq (h_{c})_{i_{n}} $. Then GWRP is defined as 

\begin{equation} \label{eq1}
g(h_{c}) = \frac{1}{Z(d_{c})} \sum_{j=1}^{N} (d_{c})^{j-1} (h_{c})_{i_{j}}
\end{equation}

\noindent where $ 0 \leq d_{c} \leq 1 $ is a hyper parameter and $ N=TF $ is the number of T-F units in a segmentation mask and $ Z(d_{c}) = \sum_{j=1}^{N} (d_{c})^{j-1} $ is a normalization term. When $ d_{c}=0 $ and $ d_{c}=1 $, GWRP simplifies to GMP and GAP, respectively. 

\subsection{Sound event detection}
The segmentation masks obtained from the JSC model contains the presence of the audio events in a T-F representation (Fig 3). Hence we achieved sound event segmentation in T-F domain. In this paper we simply average out the frequency axis to obtain the SED in the time axis. 

\section{Experiments}
In this section we apply the proposed JSC model on the modified detection of rare audio sound events dataset from Task 2 of DCASE 2017 challenge \cite{mesaros2017dcase}. This dataset consists of rare events including ``babycry'', ``gunshot'' and ``glassbreak''. The background sounds come from the acoustic scene dataset from Task 1 of the DCASE 2017 data challenge \cite{mesaros2017dcase}. To investigate WLD, we extract several rare audio events from the dataset and mix the rare audio events with 4 second clips from the acoustic scene dataset. Altogether 1008 clips are created for training, with 1/3 are single labelled and 2/3 are multilabelled. Only the presence or absence of the audio events in an audio clip is known. The audio mixtures are converted to monaural, and the sampling rate is 16 kHz. A log Mel spectrograms with 64 frequency bins are used as the T-F representation. In the Fourier transform a Hamming window with size of 1024 and overlap of 280 samples is used to ensure that there are 128 frames in each 4 seconds clip. We apply a Visual Geometry Group \cite{simonyan2014very} like CNN consists of 8 convolutional layers. Each layer consists of 64 feature maps followed by batch normalization (BN) \cite{ioffe2015batch} and ReLU nonlinearity. Dropout rate of 0.3 is applied to regularize overfitting. The value of $ d_{c} $ in GWRP is set as 0.999. These hyper-parameters are chosen empirically, but they do not affect the result much.

\begin{figure*}[t]
  \centering
  \centerline{\includegraphics[width=0.95\textwidth]{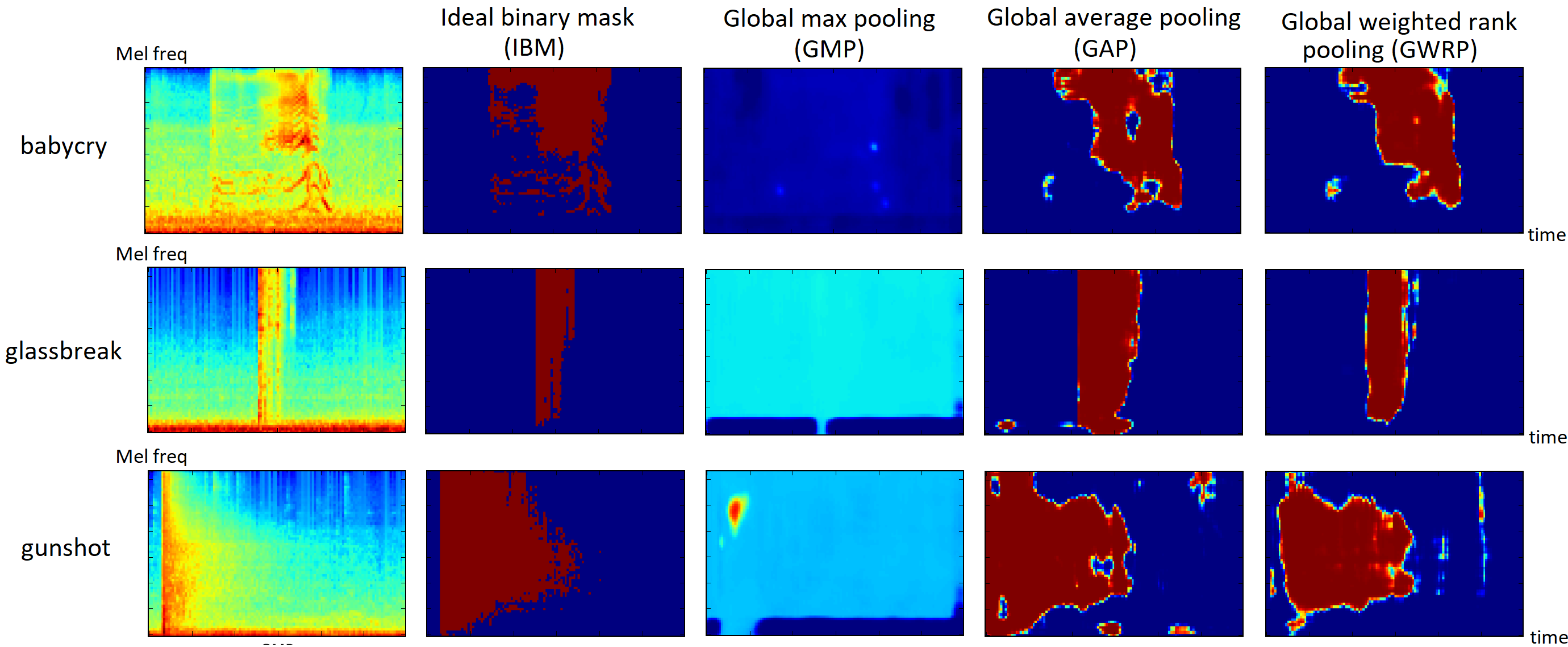}}
  \caption{Visualization of the segmentation masks using different global pooling strategy. The first column shows the log Mel spectrogram of babycry, glassbreak and gunshot sound in noisy background. The second column shows the ideal binary mask. The third to the fifth column shows the T-F segmentation masks learned using global max pooling (GMP), global average pooling (GAP) and global weighted rank pooling (GWRP), respectively. }
  \label{fig:flowchart}
\end{figure*}

\begin{table*}[t!]
	\centering
	
	\caption{Separation results of mixed rare events with background sound using different methods. }
	\resizebox{\textwidth}{!}{%
	\begin{tabular}{ p{3cm} p{1cm} p{1cm} p{1cm} p{1cm} p{1cm} p{1cm} p{1cm} p{1cm} p{1cm} p{1cm} p{1cm} p{1cm}}
		\hline 
		& Babycry & & & Glassbreak & & & Gunshot & & & Avg. \\
		\hline 
		& SDR & SIR & SAR & SDR & SIR & SAR & SDR & SIR & SAR & SDR & SIR & SAR \\
		\hline
		w/o separation & -3.66 & -3.66 & inf & -7.52 & -7.52 & inf & -6.48 & -6.48 & inf & -5.89 & -5.89 & inf \\
		IBM & 20.14 & 34.73 & 20.32 & 18.62 & 37.35 & 18.70 & 15.24 & 33.04 & 15.35 & 18.00 & 35.04 & 18.12 \\
		\hline
		Proposed GMP & 2.99 & 15.43 & 5.85 & -1.79 & 0.79 & 10.05 & -1.11 & 1.66 & \textbf{9.84} & 0.03 & 5.96 & 8.58 \\
		Proposed GAP & 9.58 & \textbf{22.61} & 10.21 & 6.35 & 17.81 & 8.49 & \textbf{2.25} & 13.05 & 4.73 & 6.06 & 17.82 & 7.81 \\
		Proposed GWRP & \textbf{13.36} & 24.61 & \textbf{14.20} & \textbf{12.29} & \textbf{28.06} & \textbf{12.86} & -1.41 & \textbf{13.93} & -0.28 & \textbf{8.08} & \textbf{22.20} & \textbf{8.93}\\
		\hline
	\end{tabular}}
\end{table*}

\begin{table}[h]
\centering
\caption{Frame wise equal error rate (EER) of mixed rare events with background sound using different method. }
\begin{tabular}{ p{2cm} p{1.1cm} p{1.2cm} p{1.1cm} p{1cm} }
 \hline 
 & babycry & glassbreak & gunshot & avg. \\
 \hline
 baseline DNN & 0.27 & 0.26 & 0.34 & 0.29 \\ 
 weak GMP & 0.27 & 0.30 & 0.32 & 0.30 \\
 weak GAP & \textbf{0.11} & 0.12 & \textbf{0.19} & \textbf{0.14} \\
 weak GWRP & \textbf{0.11} & \textbf{0.10} & 0.20 & \textbf{0.14} \\
 \hline
\end{tabular}
\end{table}

The learned segmentation masks using different classification mappings are visualized in Fig 3. The first column shows the log Mel spectrogram of a ``babycry'', a ``glassbreak'' and a ``gunshot''. The second column shows the ideal binary mask (IBM) \cite{wang2005ideal} of the audio events. Column 3 to 5 shows the segmentation masks learned using GMP, GAP and GWRP as classification mapping, respectively. It can be observed that GMP tends to underestimate the presence of the audio events in the T-F segmentation mask. GAP and GWRP performs better in learning the T-F segmentation mask on this dataset. 

Table 1 shows the separation results of different audio tags evaluated on SDR, SIR and SAR \cite{vincent2006performance}. The results of IBM \cite{wang2005ideal} and without separation are listed as baselines. GWRP performs better in terms of SDR and SAR in babycry, glassbreak, gunshot and background than without separation, GMP and GAP. Table 1 shows that source separation using the proposed JSC model outperforms significantly the baseline without separation. Table 1 also shows how far JSC is from the IBM in source separation. 

Table 2 shows the frame wise sound event detection equal error rate (ERR) using different global pooling strategies. GAP and GWRP outperforms the baselines DNN and GMP. The results are correspondent to the visualization of segmentation masks in Fig 3. and Table 1. We published source code of our work\footnote{https://github.com/qiuqiangkong/ICASSP2018\textunderscore joint\textunderscore separation\textunderscore classification}.

\section{Conclusion}
In this paper a joint separation-classification (JSC) model has been presented for sound event detection and source separation. A separation mapping from the input time-frequency representation to the segmentation masks and a classification mapping from each segmentation mask to each audio tag are proposed. We obtain frame wise sound event detection EER of 0.14, which outperforms the DNN baseline, and average source separation SDR of 8.08 using global weighted rank pooling compared to SDR of 0.03 using global max pooling. In future, we will research more on improving the source separation quality using the JSC model. 

\section{Acknowledgement}
This research is supported by EPSRC grant EP/N014111/1 ``Making Sense of Sounds'' and research scholarship from the China Scholarship Council (CSC). Thank to Sacha Krstulovic, Giacomo Ferroni and Adrian Stepien from Audio Analytic Ltd for discussions on audio event detection. 

\vfill\pagebreak

\bibliographystyle{IEEEbib}
\bibliography{strings,refs}

\end{document}